\documentclass[twocolumn]{aastex6}
\usepackage{textcomp}
\usepackage{amsmath}
\usepackage{graphicx}
\usepackage{caption}
\usepackage{subfigure}

\def\beq{\begin{eqnarray}}
\def\eeq{\end{eqnarray}}

\begin{document}

\title{Comparison of gravitational waves from central engines of gamma-ray bursts: neutrino-dominated accretion flows, Blandford-Znajek mechanisms, and millisecond magnetars}

\author{Tong Liu, Chao-Yang Lin, Cui-Ying Song, and Ang Li}
\affiliation{Department of Astronomy, Xiamen University, Xiamen, Fujian 361005, China}
\email{tongliu@xmu.edu.cn}

\begin{abstract}
Neutrino-dominated accretion flow (NDAF) around a rotating stellar-mass black hole (BH) is one of the plausible candidates for the central engines of gamma-ray bursts (GRBs). Because the time-variant and anisotropic emission of neutrinos from NDAFs leads to GRB variability, NDAFs can be regarded as the sources of the strong gravitational waves (GWs). We calculate the dependences of the GW strains on both the BH spin and the accretion rate. We demonstrate that for typical GRBs with either single pulse or multiple pulses, the GWs from NDAFs might be detected at a distance of $\sim 100$ kpc/$\sim 1$ Mpc by advanced LIGO/Einstein Telescope with a typical frequency of $\sim 10-100$ Hz. Besides NDAFs, the other two competitive candidates for GRB central engine are Blandford-Znajek (BZ) mechanism and millisecond magnetars. We explore the GW signals from these two as well, and compare the corresponding results with NDAFs'. We find that for a certain GRB, the possible detected distance from NDAFs is about two orders of magnitude higher than that from BZ mechanism, but at least two orders of magnitude lower than that from magnetars. The typical GW frequency for BZ mechanism is the same with that of NDAFs, $\sim 10-100$ Hz, while the typical frequency for magnetars is $\sim 2000$ Hz. Therefore, the GWs released by the central engines of adjacent GRBs might be used to determine whether there is an NDAF, a BZ jet or a magnetar in GRB center.
\end{abstract}


\section{Introduction}

About fifty years ago gamma-ray bursts (GRBs) were discovered. By the characteristic duration, they can be grouped into two classes \citep{Kouveliotou1993}. Long-duration GRBs (LGRBs) are usually regarded to be originated from the collapse of a massive star, while short-duration GRBs (SGRBs) are related to the merger events of black hole (BH)-neutron star (NS) or NS-NS binaries. In either case, the central engine of GRBs is likely to be a BH hyperaccretion system \citep[see reviews by][]{Liu2017} or a massive millisecond magnetar \citep[or protomagnetar, e.g.,][]{Duncan1992,Usov1992,Dai1998,Zhang2001,Dai2006,Metzger2011}.

In the BH hyperaccretion scenarios, if the accretion rate is very high ($\sim 0.001-10 ~M_{\odot}~\rm s^{-1}$), the photons cannot escape from the accretion disk, and only neutrinos are emitted from the disk surface. These neutrinos annihilate in the space outside of the disk and then form the primordial fireball to power a GRB. This kind of accretion disk is the so-called neutrino-dominated accretion flow (NDAF). In the past decades, accumulated studies have been done on NDAFs. Specifically, their structures, components and luminosities have been explored in great details~\citep[e.g.,][]{Popham1999,Narayan2001,Kohri2002,Gu2006,Liu2007,Kawanaka2007,Janiuk2007,Xue2013}. The NDAF has also been used to explain some phenomena related to the central engines of GRBs \citep[e.g.,][]{Liu2008,Liu2010,Liu2012,Liu2014,Liu2015a,Liu2016b,Kawanaka2012,Luo2013,Cao2014,Lin2016,Yi2017}. In particular, the detectabilities of gravitational waves (GWs) and MeV neutrinos released by NDAF as well as the possible existence of NDAFs in GRB centers have been discussed \citep[e.g.,][]{Reynoso2006,Lei2007,Sun2012,Liu2016c}. Besides NDAFs, in the BH hyperaccretion processes, the rotational energy of a BH can be efficiently extracted to power a Poynting jet via a large-scale poloidal magnetic field threading the BH horizon \citep{Blandford1977} to power GRBs \citep[e.g.,][]{Lee2000a,Lee2000b}. We call this Blandford-Znajek (BZ) mechanism afterwards.

In the age of BATSE, the quasi-periodic variability of GRBs is generally thought to be caused by the precession of jets \citep[e.g.,][]{Blackman1996,PortegiesZwart1999,Putten2003,Reynoso2006,Lei2007}. \cite{Liu2010} investigated that the jet precession driven by an NDAF around a spinning BH. The outer disk forces the BH to precess while the inner disk is aligned with the BH spin axis. Thus the total effect is that a precessed jet is feasible to the central engine of a GRB. The different lightcurve forms and spectral evolutions of GRBs may both be attributed to the different viewing effect. This jet precession model was successfully used to explain the variability of the giant X-ray bump in GRB 121027A \citep{Hou2014a} and the time evolution of the flares in GRB 130925A \citep{Hou2014b}. Subsequently, \cite{Sun2012} studied the GWs from the precession systems, and found that they could be detected by DECIGO/BBO in $\sim$ 10 Hz if GRBs occur in the Local Group ($\lesssim$ 1 Mpc).

\citet{Epstein1978} derived formulae for the GWs released from a small source due to the anisotropic axisymmetric emission of neutrinos. He found that the GWs may be generated from the anisotropic emission of neutrinos from supernovae (SNe), whose amplitudes and energies can be comparable to those from the collapsed SN cores. Then, \cite{Suwa2009} investigated this kind of neutrino-induced GWs from a BH hyperaccretion system, and concluded that they could be detected at $\sim$ 10 Mpc by DECIGO/BBO. Unfortunately, they simplified NDAFs as thin disks or oblate spheroids and also ignored the dominant factors from the dynamic characteristics of the NDAF.

In the GRB framework, regardless of the central engine type, there exists another type of GW sources, i.e., the hidden jets. \citet{Sago2004} analyzed the GWs from the internal shock in the GRB jets. Since the typical frequency is $\sim$ 0.1 Hz and the GW amplitude is $\sim 10^{-22}$, DECIGO/BBO might be able to detect such an event when the GRBs occur in the Local Group. The GWs from the decelerating phases of the GRB jets were studied \citep{Akiba2013} as well, and their typical frequency is $\sim 10-1000$ Hz. However, the characteristic amplitude is too low to be detected. The GWs radiated from accelerating uniform or structured jets of GRBs were also presented \citep{Birnholtz2013}. In addition, \citet{Hiramatsu2005} investigated the GWs with ``memory effect'' from the neutrino-driven jets in GRBs. They concluded that the GWs could be detected by ultimate-DECIGO in low frequency of $\sim 0.1-1$ Hz for LGRB cases.

Overall, there are various origins of GWs related to GRBs, including BH-NS or NS-NS mergers, collapses of massive stars, SNe, GRB central engines, and GRB jets \citep[see reviews by][]{Cutler2002,Postnov2014,Liu2017}. By studying these GWs sources and their electromagnetic counterparts, one may deeply reveal the nature of GRBs.

Up to now, several GW events from two merging massive BHs have been discovered by the advanced LIGO \citep[aLIGO,][]{Abbott2016a,Abbott2016b,Abbott2017}. And the \emph{Fermi}/GBM recorded a suspected SGRB 0.4 s after GW 150914 \citep{Connaughton2016}, which has been theoretically-modelled in many literatures \citep[e.g.,][]{Li2016,Loeb2016,Liu2016a,Zhang2016,Perna2016,Woosley2016,Zhang2016a,Janiuk2017}. Essentially, no more than two scenarios were proposed, i.e., BH hyperaccretion and charged BHs. The possible GW-GRB association and its theoretical explanations are still quite controversial. The investigation of the GW sources and their electromagnetic counterparts related to the compact objects is nowadays one of the most popular astrophysical topics. For the current GW detectors, the GWs from the compact binary mergers are the main goals. We here consider another potential candidates for detectors, which are from GRB central engines after merger events. For this purpose, in the present paper, the GWs from NDAFs and other candidates of GRB central engines are further revisited and compared.

The paper is organized as follows. In Section 2 we describe the numerical methods and main results of the GWs from NDAFs. In Section 3 we present the comparisons of GW detectabilities of three central engine models by aLIGO and Einstein Telescope (ET). Summary is done in Section 4.

\section{GWs from NDAFs}

\subsection{Model}

\citet{Xue2013} computed the one-dimensional steady global solutions of NDAFs in Kerr metric \citep[e.g.,][]{Kato2008}, incorporating detailed neutrino physics, chemical potentials equilibrium, photodisintegration, neutrino trapping, nuclear statistical equilibrium, etc. Based on 16 solutions with different accretion rates and BH spins, time-independent analytical formula were fitted, for the neutrino luminosity $\bar{L}_{\nu}$ and neutrino annihilation luminosity $\bar{L}_{\nu \bar{\nu}}$:
\beq \label{eq:lv}
\log \bar{L}_{\nu} ~\textrm{(erg s} ^{-1} \textrm{)} \approx 52.5+1.17a_* +1.17\log \dot{m},
\eeq
\beq \label{eq:lvv}
\log \bar{L}_{\nu \bar{\nu}}~({\rm{erg\ s^{-1}}})\approx 49.5+2.45a_*+2.17\log\dot{m},
\eeq
where $a_*$ ($0\leq a_* \leq 1$) and $\dot{m}$ are the mean dimensionless BH spin parameter and dimensionless accretion rate. Here $\dot{m}=\dot{M}/M_{\odot}~ \rm s^{-1}$, and $\dot{M}$ is the mass accretion rate. The formula is applicable for the accretion rate in the range of $0.01 < \dot{m} < 10$.

Actually, both the BH spin and accretion rate, even the structure and components of the disk, are in violent evolution in the activity timescale of the central engine, corresponding to the complicated GRB variability. The time evolution of the neutrino luminosity $L_{\nu} (t)$ can be structured as \citep[e.g.,][]{Suwa2009}
\beq \label{eq:lvtsingle}
L_{\nu} (t)= \bar{L}_{\nu} \Theta (t) \Theta (T-t),
\eeq
where $T$ is the activity duration of the GRB central engine and $\Theta$ is the Heaviside step function. This is for GRBs with single pulse. However, the observed complex variability of GRBs may be related directly to the underlying accretion behavior, and the intermittent time variability of the central engine should be taken into account, therefore it may be more realistic to consider the case of multiple pulses, i.e.,
\beq \label{eq:lvtmultiple}
L_{\nu}(t)=\sum_{i=1}^{N} \frac{\bar{L}_{\nu}T}{N\delta t} \Theta (t-\frac{i}{N} T) \Theta (\frac{i}{N} T+\delta t-t),
\eeq
where $N$ is the number of the pulses and $\delta t$ is the duration of one pulse. $N\delta t$ should be shorter than $T$ unless $N=1$ for the single pulse.

After an inverse Fourier transform, $L_{\nu} (t)$ can be written as
\beq \label{eq:lvtsinglefourier}
L_{\nu} (t)= \int_{-\infty}^{\infty} \tilde{L}_{\nu} (f)e^{-2\pi ift}df,
\eeq
where $f$ is the frequency.

We consider that the GRB variabilities are originated from the time-variant and anisotropic neutrino emission of NDAFs, which is resulted from the characteristics structure and the variation dynamics of NDAFs, hence the GW emissions from the hyperaccretion systems are related to the neutrino luminosity, and the typical GW frequencies correspond to the GRB variabilities.

The local energy flux of GWs can be written as \citep[e.g.,][]{Suwa2009}
\beq
\frac{dE_{\rm GW}}{D^2 d\Omega dt}= \frac{c^3}{16 \pi G} | \frac{d}{dt} h_+(t)|^2,
\eeq
where $D$ means the distance of a GRB, $\Omega$ is the solid angle, and the non-vanishing GW amplitude of NDAFs $h_+(t)$ can be estimated by \citep[for details, see e.g.,][]{Muller1997}
\beq
h_+(t)=\frac{2 G}{3 D c^4} \int^{t-D/c}_{-\infty} L_\nu (t')d t'.
\eeq
The total GW energy can be obtained as
\beq
E_{\rm GW} = \frac{\beta G}{9 c^5} \int^\infty_{-\infty} L_\nu^2 (t)d t,
\eeq
where $\beta \sim 0.47039$. By combining with Equation (\ref{eq:lvtsinglefourier}), one can deduce the GW energy spectrum as
\beq \label{eq:degwf}
\frac{dE_{\rm GW}(f)}{df} =\frac{2 \beta G}{9 c^5} |\tilde{L}_{\nu} (f)|^2.
\eeq
The characteristic GW strain can be expressed as \citep{Flanagan1998}
\beq \label{eq:hcf}
h_c (f)=\sqrt{\frac{2}{\pi ^2} \frac{G}{c^3} \frac{1}{D^2} \frac{dE_{\rm GW}(f)}{df}},
\eeq
From the above Equations, we can obtain the relations between $h_c (f)$ and $f$ for GRBs with single pulse or multiple pulses.

Moreover, the signal-to-noise ratios (SNRs) obtained from matched filtering for GW detectors can be calculated. Considering the relative orientation of a source and detector, the optimal SNR is
\beq
{\rm SNR}^2=\int^\infty_0 \frac{h_c^2(f)}{h_n^2(f)} \frac{df}{f},
\eeq
where $h_n(f) = [5f S_n(f)]^{1/2}$ is the noise amplitude and $S_n(f)$ is the power spectral density of the strain noise in the detector at frequency $f$.

\subsection{Results}

\begin{figure}
\centering
\includegraphics[scale=0.4]{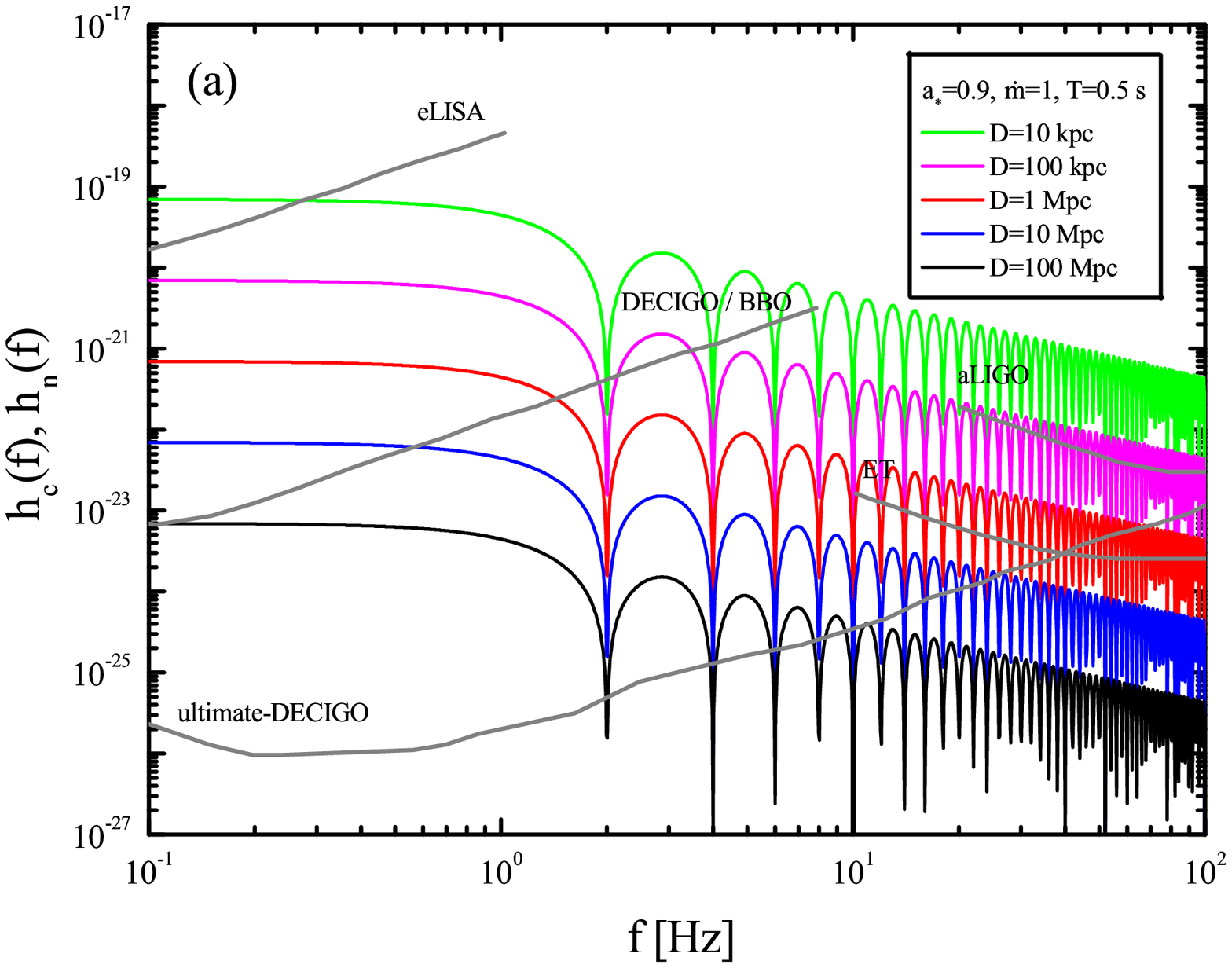}
\includegraphics[scale=0.4]{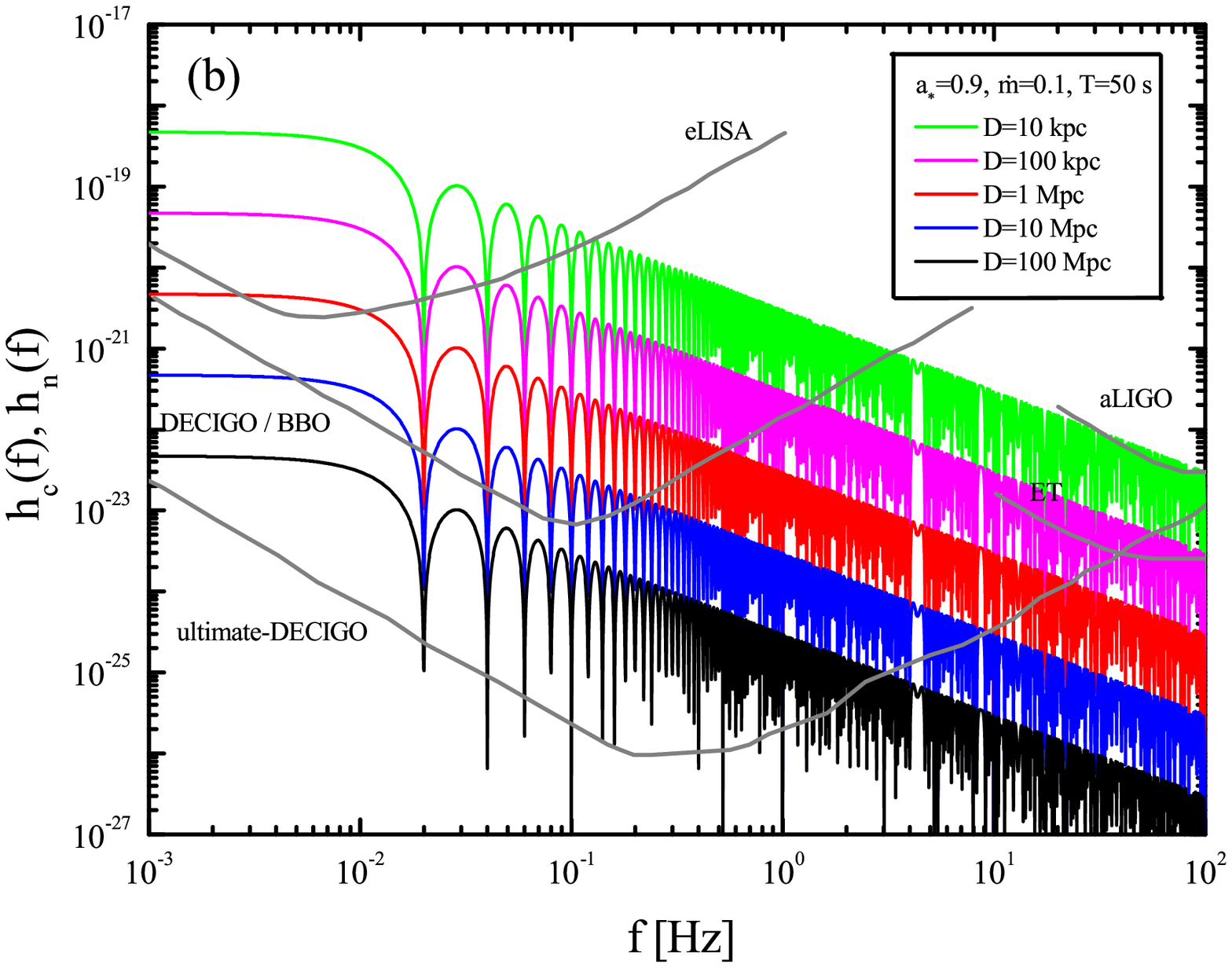}
\includegraphics[scale=0.4]{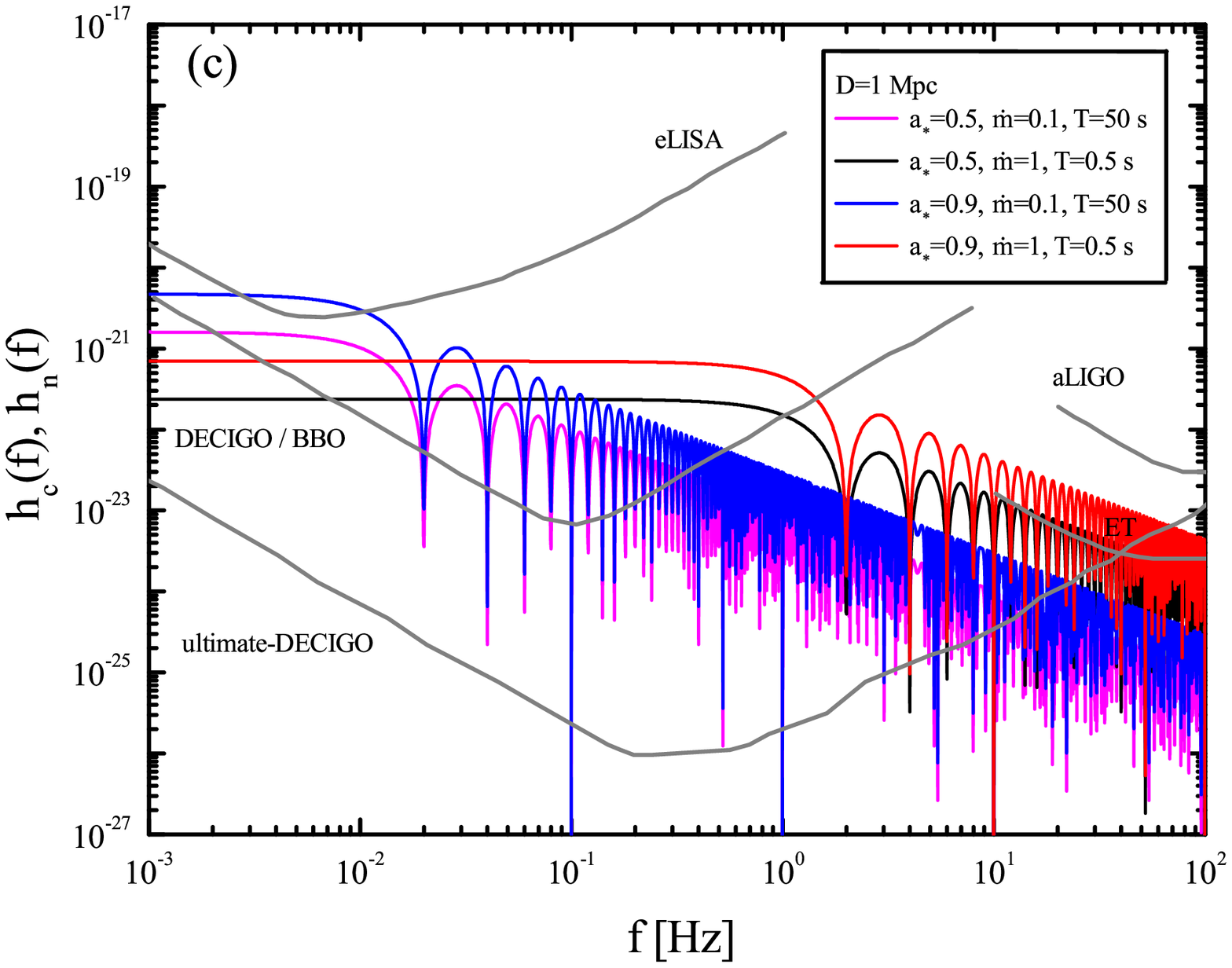}
\caption{The strains of GWs from NDAFs as the central engine of GRBs with single pulse. (a) The green, magenta, red, blue, and black lines indicate the GW strains of NDAFs with 10 kpc, 100 kpc, 1 Mpc, 10 Mpc, and 100 Mpc, respectively, for $a_* =0.9$, $\dot{m}=1$, and $T=0.5$ s. (b) Similar as (a) except $\dot{m}=0.1$, and $T=50$ s. (c) The red, blue, black, and magenta lines indicate the GW strains of NDAFs with 1 Mpc for ($a_*$, $\dot{m}$, $T$)= (0.9, 1, 0.5 s), (0.9, 0.1, 50 s), (0.5, 1, 0.5 s), (0.5, 0.1, 50 s). In all three figures, the gray lines display the sensitivity curves (the noise amplitudes $h_n$) of eLISA, DECIGO/BBO, ultimate-DECIGO, aLIGO, and ET.}
\label{fig1}
\end{figure}

\begin{figure}
\centering
\includegraphics[scale=0.4]{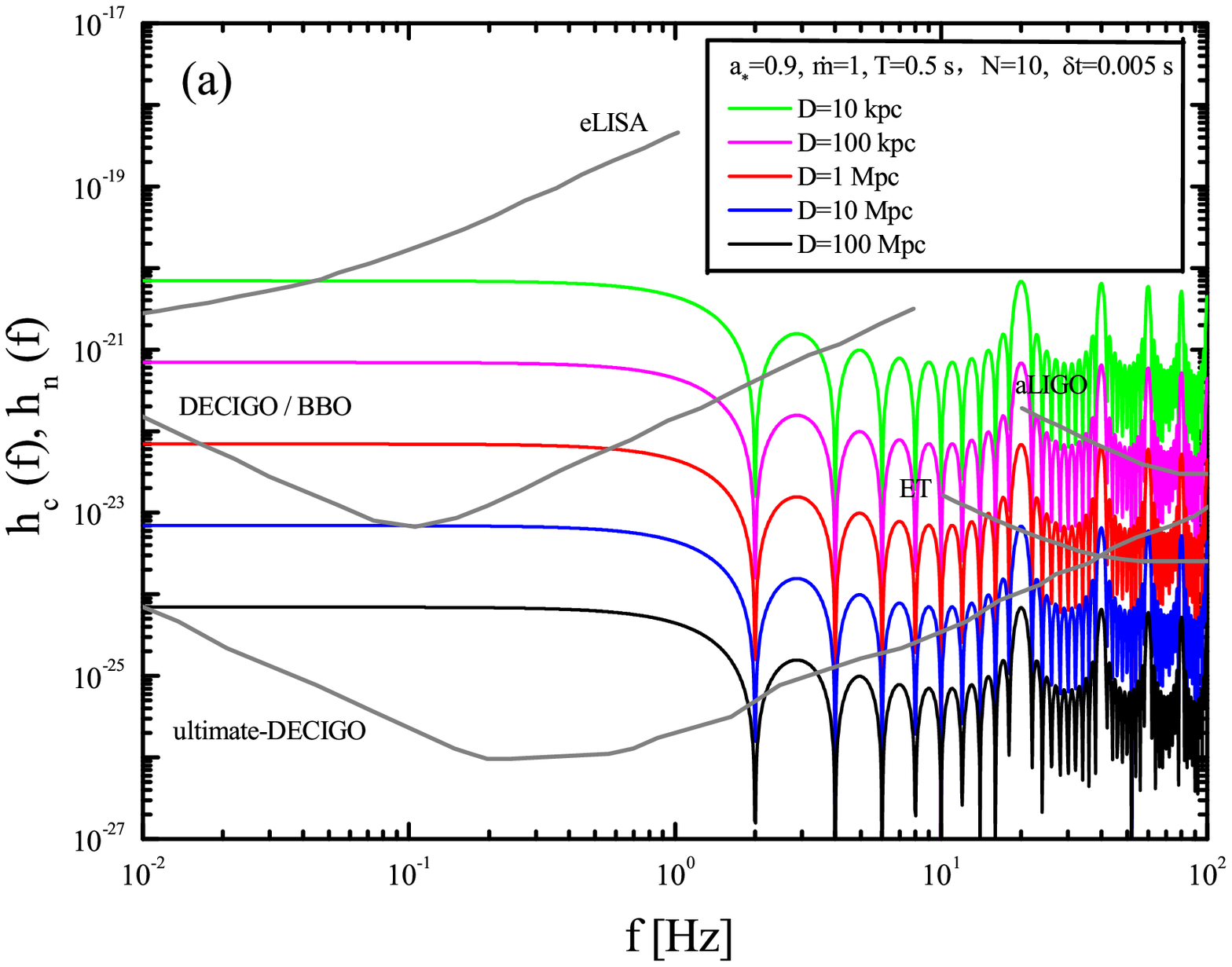}
\includegraphics[scale=0.4]{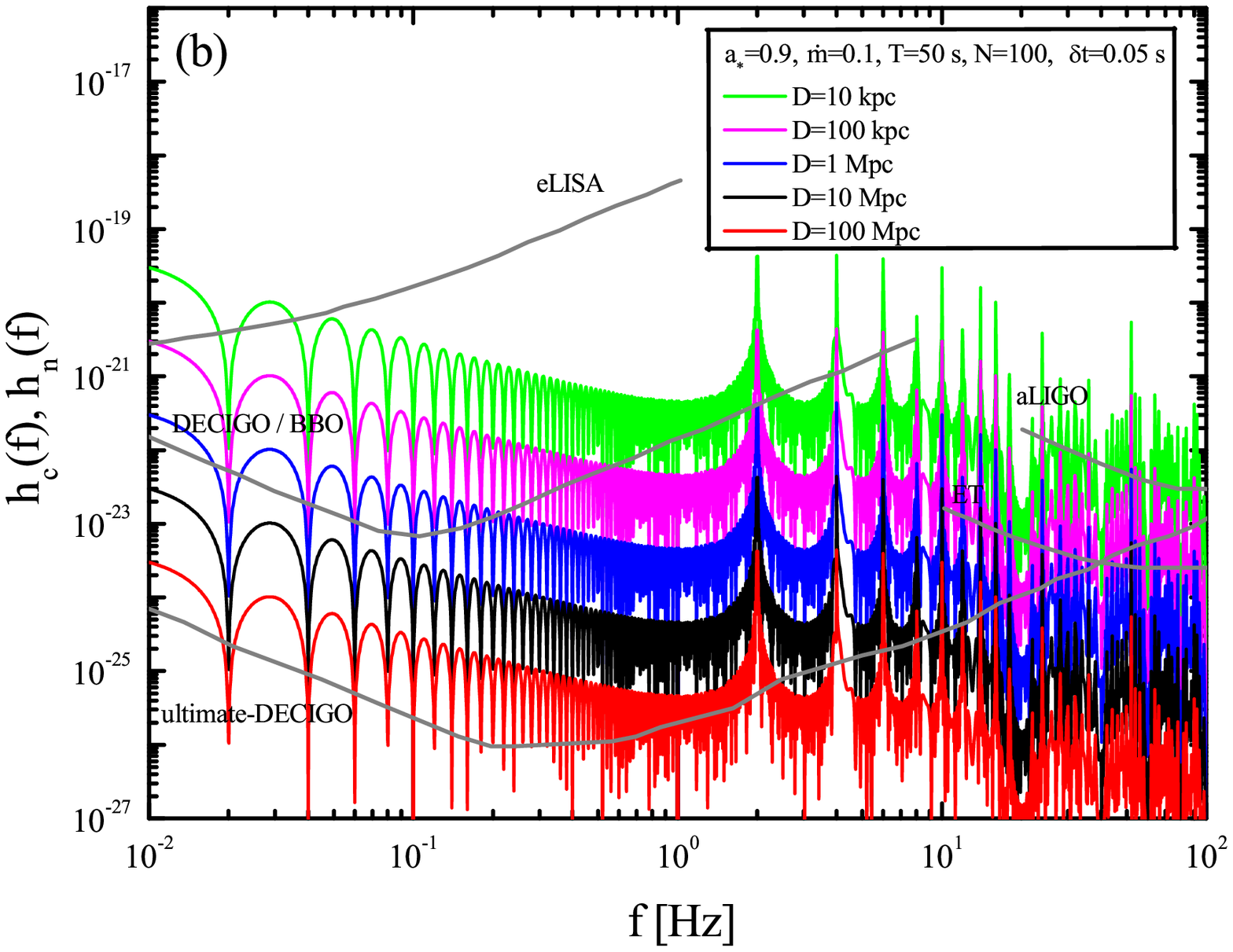}
\includegraphics[scale=0.4]{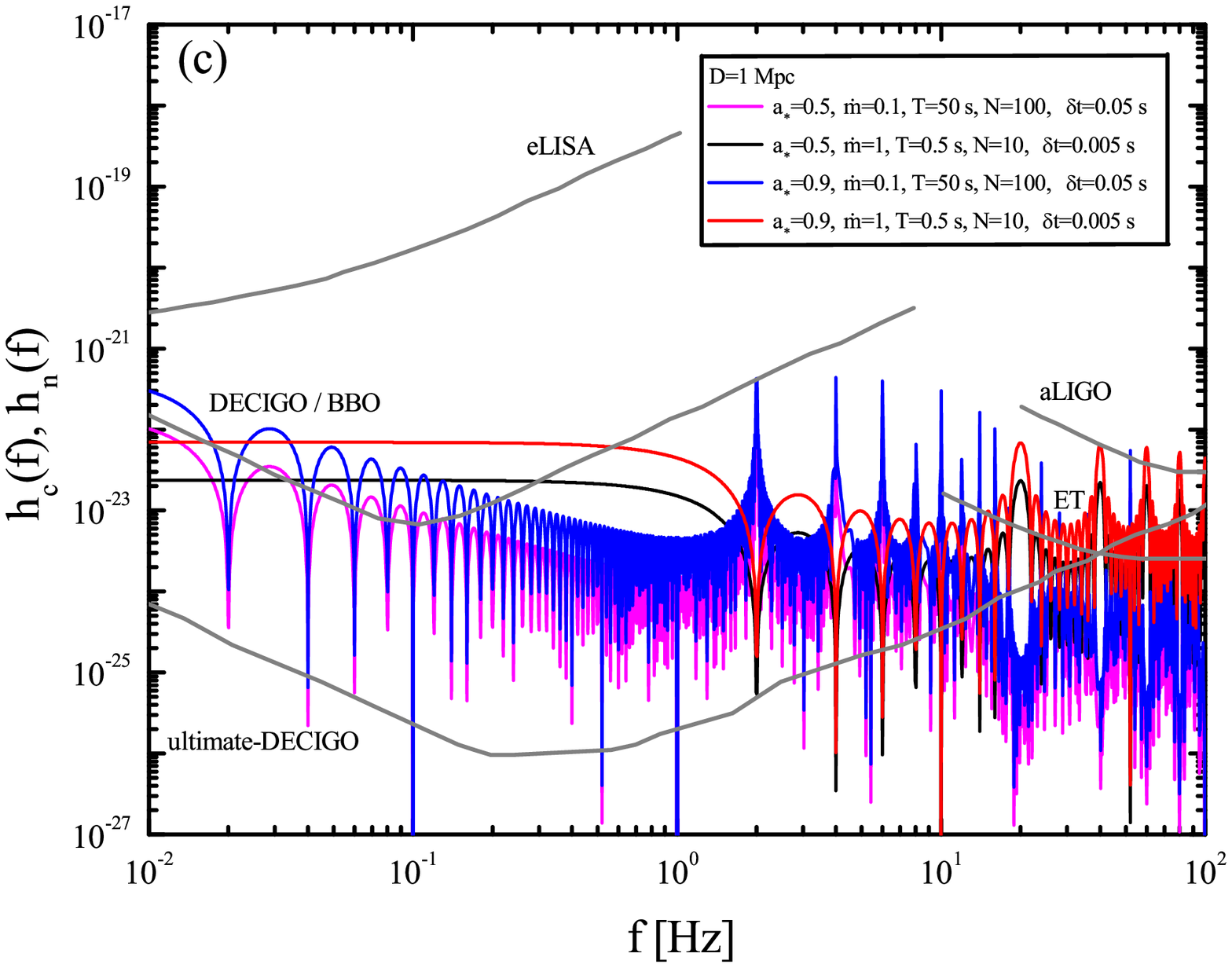}
\caption{The strains of GWs from NDAFs as the central engine of GRBs with multiple pulses. (a) The green, magenta, red, blue, and black lines indicate the GW strains of NDAFs with 10 kpc, 100 kpc, 1 Mpc, 10 Mpc, and 100 Mpc, respectively, for $a_* =0.9$, $\dot{m}=1$, $T=0.5$ s, $N=10$, and $\delta t=0.005$ s. (b) Similar as (a) except $\dot{m}=0.1$, $T=50$ s, $N=100$, and $\delta t=0.05$ s. (c) The red, blue, black, and magenta lines indicate the GW strains of NDAFs with 1 Mpc for ($a_*$, $\dot{m}$, $T$, $N$, $\delta t$)= (0.9, 1, 0.5 s, 10, 0.005 s), (0.9, 0.1, 50 s, 100, 0.05 s), (0.5, 1, 0.5 s, 10, 0.005 s), (0.5, 0.1, 50 s, 100, 0.05 s). In all three figures, the gray lines display the sensitivity curves (the noise amplitudes $h_n$) of eLISA, DECIGO/BBO, ultimate-DECIGO, aLIGO, and ET.}
\label{fig2}
\end{figure}

Before the central BHs or magnetars are born to power GRBs, the compact binary mergers and collapsars are also important GW sources \citep[e.g.,][]{Cutler2002,Postnov2014,Liu2017}. We here restrict ourselves only on the GWs from the GRB central engines.

In the NDAF model, we adopt $\dot{m}$ and $T$ as (0.1, 50 s) and (1, 0.5 s) as the typical luminosities and durations of LGRBs and SGRBs, respectively. Our main interest is the effects of the distance of GRBs to the Earth $D$, BH spin parameter $a_*$, and the mean accretion rate $\dot{m}$, on the GW strains.

Figures~\ref{fig1} and ~\ref{fig2} show the strains of GWs from NDAFs as the central engine of GRBs with single pulse and multiple pulses, respectively. It is obvious that the GW strains have positive correlations with both the BH spin parameters and accretion rates [as seen from Equation (\ref{eq:lv})], and have negative correlations with the distances of the sources as seen from Equation (\ref{eq:hcf}), i.e., $h_c$ is in the inverse proportion of $D$]. One can roughly read the SNR from these figures. Furthermore, by comparing Figures 1 and 2, we notice that for the same $T$, the spectra of GRBs for single pulse and multiple pulses are very different in the high-frequency range but similar in the low-frequency range. This is because that in the case of the multiple pulses many pulses are in short time scale and long-term behaviors are independent to the particulars of the burst.

In all three figures,the gray lines display the sensitivity curves (the noise amplitudes $h_n$) of eLISA, DECIGO/BBO, ultimate-DECIGO, aLIGO, and ET, respectively. For GRBs with either single pulse or multiple pulses, the GWs from NDAFs can be detected at a distance of $\sim 100$ kpc by aLIGO, and $\sim 1$ Mpc by ET in the detectable frequency $\sim 10-100$ Hz. they can be detected by ultimate-DECIGO at $\sim$ 100 Mpc in $\sim 0.1-10$ Hz for the LGRB cases. Additionally, in the jet precession model \citep{Sun2012}, the typical GW frequency is determined by the precession period, which corresponds to the GRB variability. For the neutrino-induced GWs from NDAFs, the GW frequency also depends on the GRB variability, which corresponds to $\sim 10-100$ Hz for aLIGO/ET and $\sim 0.1-10$ Hz for ultimate-DECIGO in LGRB cases.

\section{Comparisons of GWs from different central engine models}

NDAFs, BZ mechanism, and magnetars are three mainly possible candidates for the central engine of GRBs. They have different capabilities on powering GRBs in the scenarios of the collapsars or compact object mergers, different dynamics on describing GRB morphology, and different evolution, components and products for progenitors and environment of GRBs \citep[see reviews by][]{Liu2017}. They have certainly also different GW radiations.

\subsection{BZ mechanism}

For a BZ jet driven from a BH hyperaccretion disk, its luminosity can be written as \citep{Blandford1977,Lee2000a,Lee2000b}
\beq
L_{\rm BZ}=1.7\times10^{20}a_*^{2}m^{2}B_{\rm in, G}^{2}F(a_*)~{\rm erg~s^{-1}},
\eeq
where $B_{\rm in, G}=B_{\rm in}/1 \rm G$ is the dimensionless magnetic strength at the inner boundary of the disk. $m=M/M_{\odot}$, with $M$ the BH mass.
\beq
F(a_*)=[(1+q^{2})/q^{2}][(q+1/q)\arctan(q)-1],
\eeq
where $q=a_*/(1+\sqrt{1-a_*^{2}})$. According to the balance between the pressure of the disk and the magnetic pressure on the BH horizon, the BZ jet power can be derived as
\beq \label{eq:lbz}
L_{\rm BZ}=9.3\times10^{53}a_*^{2}\dot{m} F(a_*)(1+\sqrt{1-a_*^{2}})^{-2}~{\rm erg~s^{-1}}.
\eeq

Comparing Equations (\ref{eq:lvv}) to (\ref{eq:lbz}), one can see that for fixed BH spin and accretion rate, $L_{\rm BZ}$ is about two orders of magnitude larger than $\bar{L}_{\nu \bar{\nu}}$. On the other hand, if assuming that two mechanisms have the same conversion efficiency to power a certain GRB, hence $L_{\rm BZ} = \bar{L}_{\nu \bar{\nu}}$, then the BH spin and accretion rate are one or two orders of magnitude lower than those in NDAFs, respectively. This means that $\bar{L}_{\nu}$ should be about two orders of magnitude larger than $L_{\rm BZ}$, as shown in Equations (\ref{eq:lv}) and (\ref{eq:lvv}).

Since BZ mechanism and neutrino annihilation process in NDAFs both depend on BH hyperaccretion systems and their dynamical characteristics, the physical mechanisms of the GWs are therefore similar. Then from Equations (\ref{eq:degwf}) and (\ref{eq:hcf}), we know that the GW strain from BZ mechanism is lower than that from NDAFs. That is, for a certain GRB, as shown in the right panel of Figure \ref{f3}, the GW detectable distance of BZ mechanism is about two orders of magnitude lower than that of NDAFs. Furthermore, as mentioned above, the GW frequency in the BH hyperaccretion framework, no matter which mechanism, is obviously determined by the typical GRB variability, so the GW detectable frequency of BZ mechanism by aLIGO and ET is $\sim 10-100$ Hz.

\subsection{Magnetars}

\begin{figure*}
\centering
\includegraphics[scale=0.45]{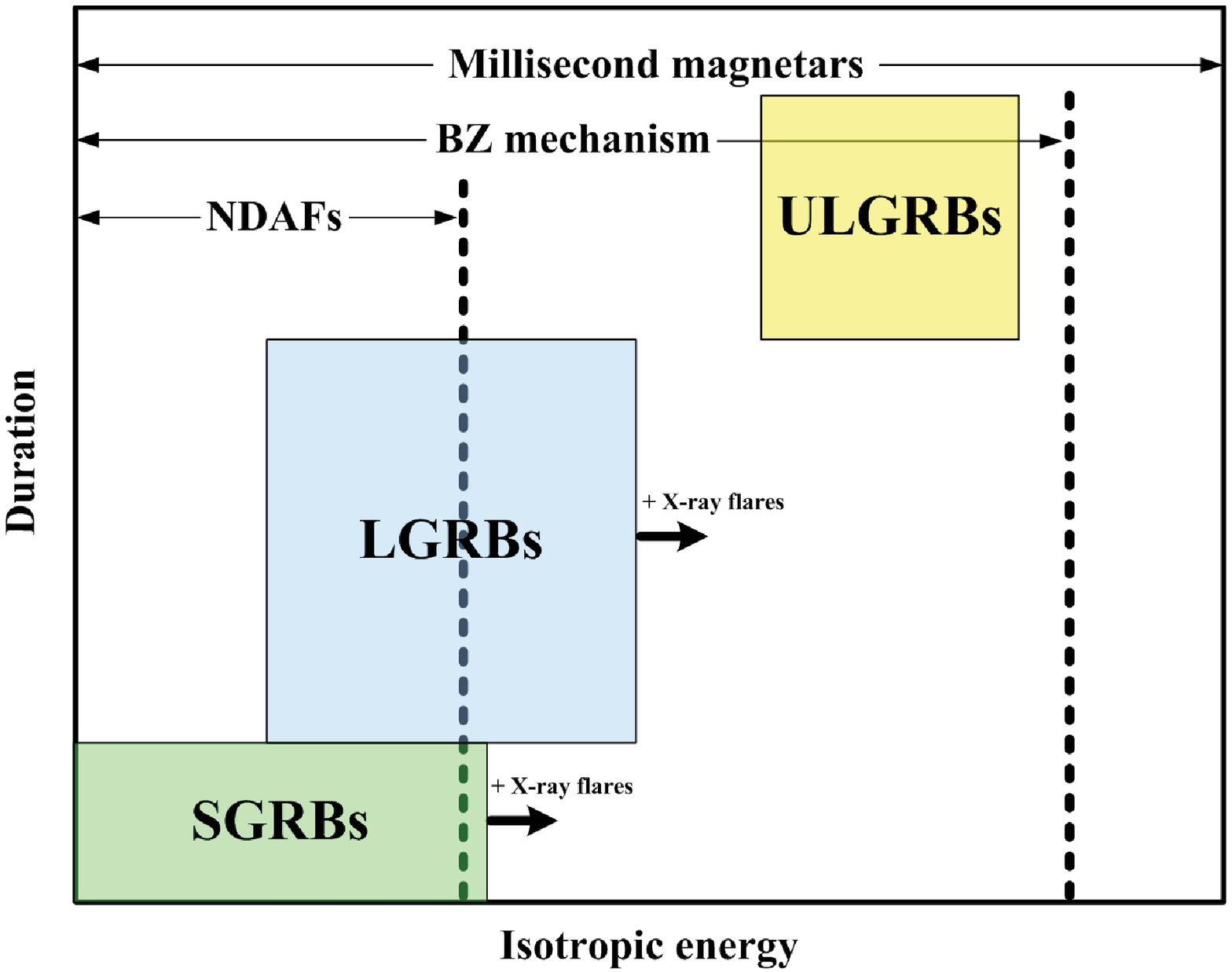}
\includegraphics[scale=0.45]{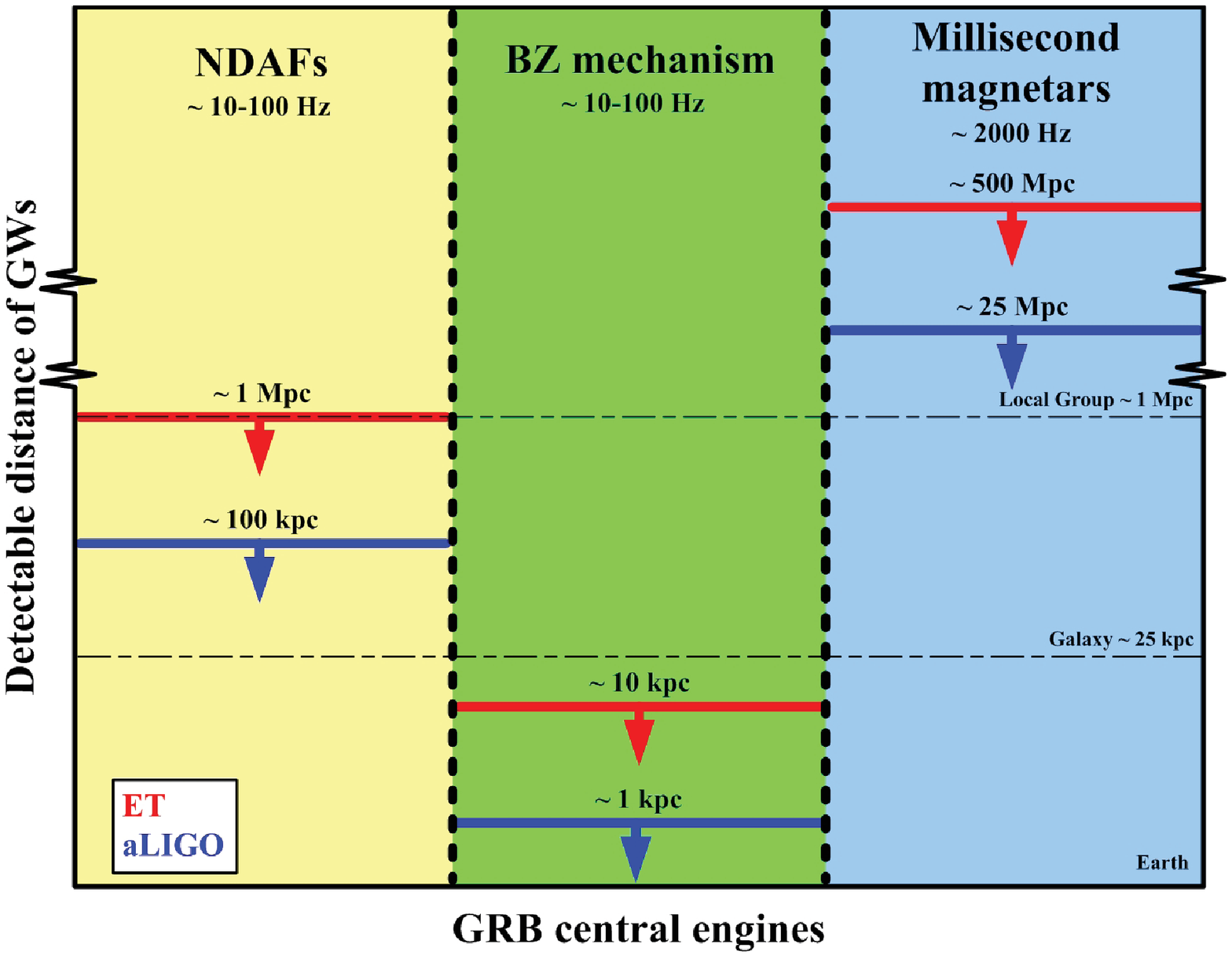}
\caption{Left panel: Schematic picture of the applications of three GRB central engine models to the isotropic energy of three types of GRBs (adapted from Figure 20 in \citet{Liu2017}). Right panel: Schematic picture of the GW detectable distances of three models by aLIGO (blue lines) and ET (red lines).}
\label{f3}
\end{figure*}

Magnetars have been widely studied as the central engine of GRBs \citep[e.g.,][]{Duncan1992,Usov1992,Dai1998,Zhang2001,Dai2006,Metzger2011}, which are described as rapidly spinning (the typical rotation period $P \sim 1$ ms), supramassive ($\sim 2.6-2.8 ~M_\odot$), and strongly magnetized (the dipole magnetic field strength $B \sim 10^{15} ~\rm G$) NSs. The spin-down of magnetars from NS-NS merger events has been used to explain some GRBs with an internal X-ray plateau \citep[e.g.,][]{Rowlinson2010,Yu2010,Lu2015,Gao2016}. Recently
the equation of state models of quark stars
have been suggested to be more preferred than those of NSs for such bursts with plateaus \citep[e.g.,][]{Lia2016,Li2017}. LGRBs and even super-luminous SNe have been investigated in the scenario of magnetar born in the collapsars \citep[e.g.,][]{Metzger2008,Metzger2011,Metzger2015}. Further studies have included also ultra-LGRBs (ULGRBs) in the magnetar scenario \citep[e.g.,][]{Greiner2015,Ioka2016}. That is, magnetars might power all types of GRBs and reconcile their diverse behaviors.

We mention here that the total rotational energy of a magnetar can be estimated as \citep[e.g.,][]{Lu2015}
\beq
E_{\rm rot}\approx2\times 10^{52} (\frac{M_{\rm NS}}{1.4~M_\odot})(\frac{R_{\rm NS}}{10^6 ~\rm cm})^2 (\frac{P}{1~\rm ms})^{-2}~\rm erg,
\eeq
where $M_{\rm NS}$ and $R_{\rm NS}$ are the mass and radius of the magnetar, respectively.

Concerning the accretion rate and timescale (i.e., the accreted mass), the BH hyperaccretion systems are more demanding as the GRB progenitors than magnetars. Furthermore, BZ mechanism can power a GRB more easily than NDAFs, as demonstrated in Section 3.1. Those are summarized in the left panel of Figure \ref{f3} (adapted from Figure 20 in \citet{Liu2017}): First, almost all SGRBs might be described in the NDAF model, with reasonable disk masses derived from the remanent of the compact objects mergers \citep{Liu2015c}; Second, only about half of LGRBs might satisfy the NDAF model, while it might be necessary for the left half to introduce massive disks, extreme Kerr BHs, and high conversion efficiency for neutrino annihilation \citep{Song2016}; Third, for LGRBs and ULGRBs, BZ mechanism is especially more efficient than NDAFs. Actually, the deviation between BZ mechanism and NDAFs is more significant if X-ray flares are included \footnote{The total energy of GRBs generally includes the isotropic radiated energy in the prompt emission phase and the isotropic kinetic energy of the jet powering long-lasting afterglow. And some flares may be origin from the restart of GRB central engine.} \citep[e.g.,][]{Liu2015b,Luo2013,Mu2016}; Fourth, the millisecond magnetar model could cover almost all types of GRBs as said before.

\citet{Cutler2002} reviewed the estimations for the event rates and the GW strengths of the well-known GW sources including NSs. The quadrupole deformation of a NS in the spin-down phase, caused by the rapid spin or magnetic field, can be described by the ellipticity $\epsilon$, which leads to the GW radiation. Formerly, the ellipticity $\epsilon$ of NSs was usually as small as $\sim 10^{-5}-10^{-6}$, which cannot result in detectable GWs \citep{Andersson2003}. Recently, the GW radiation of magnetars has been carefully revisited \citep{Corsi2009,Fan2013a,Fan2013b,Dai2016,Gao2017}, a larger $\epsilon \sim 0.005$ might be reachable. On the other hand, the initial rotation period of a newborn magnetar is expected to be $\sim$ 1 ms, whatever is originated from NS-NS mergers or collapsars. Its rotational energy might be larger than the energy requirements of GRBs and kilonovae \citep[or mergernovae, see e.g.,][]{Yu2013,Metzger2017}. The remaining energy has to be carried away by a non-electromagnetic emission, i.e., GWs \citep{Fan2013b}. Consequently, the energy of the GW radiation from magnetars approaches the typical GRB energy, which should be much higher than the BH hyperaccretion systems. In the jet precession model of magnetars \citep{Sun2012}, the quadruple power of GWs is six orders of magnitude lower than GRB luminosity at $f \sim 10$ Hz, like in the cases of NDAFs and BZ mechanism.

Following \citet{Fan2013a}, the GW radiation from millisecond magnetars has a frequency of $f \sim$ 2000 Hz, corresponding to $P \sim 1$ ms \citep[$f=2/P$, e.g.,][]{Zimmermann1979,Shapiro1983}, and the characteristic GW amplitude at this frequency can be estimated as
\beq
h_c \approx 5.1 \times 10^{-22} (\frac{D}{100~\rm Mpc})^{-1}(\frac{I}{10^{45.3} ~{\rm g ~cm^2}})^{1/2}(\frac{P}{1~\rm ms})^{-1},
\eeq
where $I$ is the moment of inertia of the NS. From this equation, we can estimate the GW detectable distance of a typical GRB originated from magnetars. It is $\sim$ 25 Mpc for aLIGO and $\sim$ 500 Mpc for ET, which is consistent with the estimations in \citet{Gao2017}. As our main results, the GW detectable distances of three central engine models are displayed in the right panel of Figure \ref{f3} for a typical GRB. From the distance of the sources, the characteristic frequency, and the GW amplitude, one can determine whether an NDAF, a BZ jet or a magnetar in the GRB center.

However, the rate of GRBs occurred in these GW detectable distances is apparently low. \citet{Wanderman2015} declared that the ``local'' SGRB detection rate is about $4.1^{+2.3}_{-1.9}$ Gpc$^{-3}$ yr$^{-1}$. While, the SGRB rate is much lower than the LGRB rate \citep[e.g.,][]{Fryer1999,Podsiadlowski2004,Virgili2013}. \citet{Liu2016c} calculated the LGRB rates, including off-axis LGRBs, of the major galaxies in the Local Group (shown in Table I), which is about 3 per century. It can be considered roughly as the upper limit of GW detection rate in the Local Group.

\section{Summary}

In the lifetime of GRBs, the progenitors which include compact binary mergers and collapsars, the central engines which include BH hyperaccretion systems and millisecond magnetars, and the jets which include internal and external shocks, are all potentially detectable GW sources, and can be detected by the current or future detectors when occuring in the nearby galaxies. Furthermore, for a certain step, the differences in the strength, the typical frequency, and the detectable distance of the GWs can be used to identify the underlying mechanisms.

Also, in this paper we proposed a new way, besides the MeV neutrino emission \citep[e.g.,][]{Liu2016c}, to distinguish different GRB central engine models, i.e., NDAFs, BZ mechanism, and millisecond magnetars. It is through their GWs. For a typical GRB, the detectable distances of the three models on aLIGO/ET are roughly 100 kpc/1 Mpc at $\sim 10-100$ Hz, 1 kpc/10 kpc at $\sim 10-100$ Hz, and 25 Mpc/500 Mpc at $\sim 2000$ Hz, respectively.

It is possible to detect the GWs from NS-NS mergers under the current detectability. First, a bright GRB, an X-ray transient or an optical kilonova are likely to be observed \citep[e.g.,][]{Liu2017,Metzger2017}. Second, after merger events, a possible newborn millisecond magnetar will release strong GWs, which may be detected by aLIGO in the high-frequency range; if not, it is a BH hyperaccretion system that might exist in the GRB center.

\acknowledgments
We thank the anonymous referee for very useful suggestions and comments. We also thank Mou-Yuan Sun and Tuan Yi for helpful discussion. This work was supported by the National Basic Research Program of China (973 Program) under grant 2014CB845800 and the National Natural Science Foundation of China under grants 11473022 and U1431107.

\clearpage

\end{document}